\documentclass[runningheads]{llncs}
\usepackage{graphicx}
\usepackage{hyperref}

\begin{document}
\title{Leveraging Machine Learning for Ransomware Detection}
%
%
\author{Nanda Rani\inst{1,2}\orcidID{0000-0003-1255-5284} \and
Sunita Vikrant Dhavale\inst{1,2}\orcidID{0000-0003-0306-6198}}
\authorrunning{Nanda et al.}
%
\institute{Defence Institute of Advanced Technology, Pune, India \and
\email{\{nanda\_mcse19,sunitadhavale\}@diat.ac.in}}
\maketitle              
\begin{abstract}
The current pandemic situation has increased cyber-attacks drastically worldwide. The attackers are using malware like trojans, spyware, rootkits, worms, ransomware heavily. Ransomware is the most notorious malware, yet we didn’t have any defensive mechanism to prevent or detect a zero-day attack. Most defensive products in the industry rely on either signature-based mechanisms or traffic-based anomalies detection. Therefore, researchers are adopting machine learning and deep learning to develop a behaviour-based mechanism for detecting malware. Though we have some hybrid mechanisms that perform static and dynamic analysis of executable for detection, we haven’t any full proof detection proof of concept, which can be used to develop a full proof product specific to ransomware. In this work, we have developed a proof of concept for ransomware detection using machine learning models. We have done detailed analysis and compared efficiency between several machine learning models like decision tree, random forest, KNN, SVM, XGBoost and Logistic Regression. We obtained 98.21\% accuracy and evaluated various metrics like precision, recall, TP, TN, FP, and FN.

\end{abstract}
\section{Introduction}

The US colonial pipeline was recently attacked by ransomware named DarkSide and paid \$ 4.4 million ransom, which was requested in ransom note~\cite{article1}. The attackers stole approx. 100GB data and threatened to leak if the ransom was not being paid. Due to the attack, the country faced fuel shortages and disruption in oil-related services. Nowadays, Attackers are primarily targeting critical infrastructure to disrupt the services that can harm and affect a country population or even a human life if healthcare is being targeted. In history, ransomware has also compromised health care, which causes even human life in danger. Not only critical infrastructure but some giant companies are also being targeted by ransomware, and they are pressured to pay the ransom to get their system access and data back. Attackers generally encrypt
victim's data with their key, and they return the key or decrypt the data after victims pay them demanded ransom. Earlier, ransomware
was only disrupting the services and making the data unavailable to the victim; they were still safely stored only on their system.  But nowadays, data is also being stolen by attackers, threatening the victim to pay the ransom. Though it is not guaranteed that data can get back after paying the ransom or they didn't leak the stolen data.

Ransomware is a class of malware that encrypts the victim data and make them not accessible to the victim. An attacker generally displays a ransom note after encrypting the victim data, which usually states that the attack has been performed and demanded ransom. The steps to make the payment they also provided with that ransom note
only. Some of the attackers also create a negotiation window for ransom negotiation. Ransoms are typically demanded in bitcoins because of their anonymity, so attackers cannot get tracked from payment gateway  \cite{ref_article2}. Ransomware is categorized into two types  \cite{ref_article3}: (1) Locker ransomware, (2) Crypto ransomware.

Locker ransomware only locks the system, and the victim won’t be able to use the system. Ransom notes usually displayed by attackers consist of payment information. This attack usually happens by locking the victim's desktop or displaying a ransom note that never closes. Victim data is not compromised in this attack, and in some cases, there is a recovery option to restart the system in safe mode and perform countermeasures. Sometimes locker ransomware, also known as Screen-Locker ransomware  \cite{ref_article4}. Crypto ransomware encrypts the user files and adds their extension to the victim files, which makes them inaccessible to use. In some cases, data is also being stolen during communication between the victim system and command \& control. After encryption, a ransom note usually displays consisting of payment information and demanded ransom. Sometimes crypto-ransomware is also known as File-Locker ransomware \cite{ref_article4}. If we think from the CIA triad perspective, availability compromises in any locker ransomware attack. But in the case of crypto-ransomware, Integrity and availability compromise typically, and in some cases, confidentiality also compromises.

\begin{figure}
\includegraphics[width=\textwidth]{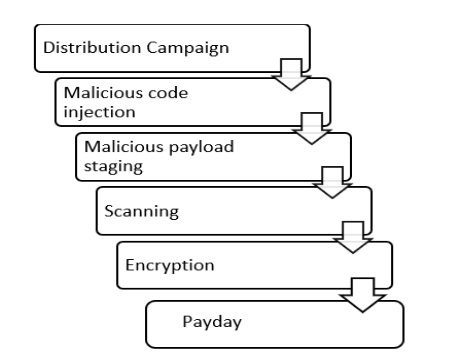}
\caption{Ransomware attack flow kill-chain stages.} \label{fig1}
\end{figure}

Numerous ransomware variants have been introduced yet, such as WannaCry \cite{ref_article9}, Cryptowall \cite{ref_article5}, Gancrab \cite{ref_article6} , REvil \cite{ref_article10}, Jigsaw \cite{ref_article11}, Ryuk \cite{ref_article12}, Petya \cite{ref_article13}, Cryptolocker \cite{ref_article7}, Teslacrypt \cite{ref_article8}. WannaCry ransomware attack was a notorious attack that occurred worldwide in 2017 and affected 150 countries \cite{ref_article14}. Many healthcare services and hospitals were affected during this attack period. Attackers have exploited the EternalBlue vulnerability in the SMB (Server Message Block) protocol of Windows frameworks \cite{ref_article15}. Wannacry, Cryptolocker ransomware came into the picture in 2013, and it was the highest ransom of its time and infected millions of windows systems \cite{ref_article9}. Like cyber kill chain process, One of the Exabeam threat research report generalized ransomware attack flow kill chain process in six stages shown in Figure \ref{fig1} \cite{ref_article16}:

\begin{enumerate}
    \item \emph{Distributed Campaign:} Victims are tricked by the attacker to download their dropper code using various techniques such as social engineering.
    \item \emph{Malicious Code Injection:} Dropper code download and ransomware executable introduce to the victim machines.
    \item \emph{Malicious Payload Staging:} Insert ransomware into the victim system and create a persistence connection with the command \& control server.
    \item \emph{Scanning:} Compromised system and network reachable resources being scanned by installed executables for encryption.
    \item \emph{Encryption:} All targeted data will be encrypted in this stage.
    \item \emph{Payday:} After encryption, a ransom note will be displayed, and data will be inaccessible to use for the victim.
\end{enumerate}

Encryption techniques used by ransomware are (1) Symmetric encryption, (2) Asymmetric encryption, and (3) Hybrid encryption. Attackers use a single key for encryption and decryption in symmetric encryption. They share a common key while communicating with the C\&C server or encode the key with dropper code. Forensic experts can extract the common key if it is encoded with dropper code. While attackers use different keys for encryption and decryption in asymmetric keys, it would be challenging to extract them using forensic methods. But communication made during key distribution between the victim system and C\&C server can be captured by monitoring traffic. But it fails when attackers use proxy servers to communicate with a compromised system. Afterwards, attackers started using a hybrid encryption technique, in which files would be encrypted by symmetric key only, but another asymmetric key will encrypt that symmetric key. So, if forensic teams extract the symmetric key, they won’t be able to use it as it is encrypted with asymmetric encryption, whose decryption is with the attacker only.

The attacker also uses various obfuscation techniques to evade themselves in the system \cite{ref_article17}. Malware tries to check inconsistencies in hardware like number of CPU cores to investigate sandbox existence. If the malware finds its execution in the sandbox environment, they hide their malicious indication or behaviours during analysis. Nowadays, most sandbox merchants try to hide their genuine configuration to trick the malware into confidence that it is being executed within the actual host. Several static analysis tools can unpack the packed malware and then unpacked malware can be used for analysis.

\section{Literature Review}

To analyze ransomware, we have different ways to evaluate any executables: (1) Static Analysis - In this method, executables are being examined without running them, and it is useful to detect malicious infrastructure, library or packed files. Technical indicators are recognized, like file name, hash, a string such as IP address, domain name, and file's header data, which can get utilized to decide whether that file is malicious or not. In addition, tools like disassemblers and network analyzers can spot the malware without actually running it to collect evidence on how that executable malware works. (2) Dynamic Analysis: - In dynamic malware analysis, it executes suspected malicious code or programs in a safe environment called a sandbox. This closed system allows security professionals to investigate the malware in action without the risk of letting it infect the targeted system or escape into the enterprise network. Cuckoo sandbox is the most useful tool when it comes to analyzing malware dynamically. (3) Hybrid Analysis: It delivers the security team the best of both approaches by merging static and dynamic analysis techniques. It can detect malicious programs that hide themselves and then extract many indicators of compromise (IOCs) by statically and previously unseen programs. It also helps to detect unknown threats, even those from the utmost sophisticated malware.

The signature-based protection works on the concept of having a signature covering the families of ransomware. We can consider a signature as a “fingerprint”. The detection system matches files signatures from the stored signature database. This system works relatively well for known malware but does not cover threats they know about, i.e., zero-day threats. Abnormal traffic detection is considered as one step up from signature-based detection. Anomalous traffic is detected based on many different parameters like network intrusion detection (NID) and other traffic detections recognized as malicious. The main shortcoming of using the abnormal traffic detection method is its high false-positive rate. Hence, there is a decent chance that genuine network traffic can get identified inaccurately. Thus, we need to monitor process activity and the host’s backend operations to decide ransomware infection presence. Due to the emerging Machine Learning technology, it's proving a vital role in monitoring endpoint properties and identifying patterns of system activities for any malware. When machine learning is implemented in file behaviour detection in any system, this can create an effective solution for detecting ransomware. Machine Learning (ML) is similar to human learning in a sense. Legitimate program execution and application present a specific type of behaviour. Over time, ML “learns” how genuine, regular programs act by taking in enormous amounts of data facts through dedicated investigation that may include interactive debugging. In this detailed and inclusive investigation of legitimate program execution, ML develops very good at recognizing programs that are impersonators and have malicious intentions.

There are different methods and techniques to detect ransomware \cite{ref_article18,ref_article19,ref_article20}. Static analysis-based methods disassemble source code without executing it. But they suffer from high false-positive rates and fails to detect obfuscated ransomware. New variants are coming up frequently, and attackers use different packing techniques to mutate their codes. Researchers are moving towards dynamic behaviour analysis-based methods that monitor the interactions of the executed code in a virtual environment for fighting these issues. However, these detection methods are slower and require considerable memory resources. The machine learning approach is best suited for the analysis of the behaviour of any process or application.

Various researchers have focused on either host-based artefact analysis or network-based artefacts analysis to develop an effective defensive mechanism. Host-based artefacts like process behaviour, file system activities, API calls, I/O request packets logs, Registry Key Operations and many more are usually considered to monitor. To monitor network-level artefacts traffic can get monitored when the compromised system communicates with the C\&C server. The network-level artefacts such as source and destination IP address, protocol type, source and destination port number, the total number of bytes per packet per conversation can be considered for analysis. In the network-based analysis, there is an assumption that malware will communicate with their C\&C server for purposes like getting an encryption key. But this technique may fail if malware doesn’t contact C\&C servers or used any local key. Also, malware might use proxy servers while contacting C\&C servers, and then the network-based analysis may not be feasible to detect. Researchers focused on process anomaly detection to enhance their detection rate. The technique discussed in the paper uses Windows API calls \cite{ref_article21,ref_article22,ref_article23}, I/O request Packets (IRP) logs, File system operations, set of operations performed per file extension, directories operations, dropped files, registry key operation, strings \cite{ref_article24,ref_article25,ref_article26,ref_article27,ref_article28} for detection. Researchers recorded artefacts like folder listing, Files written, Files Renamed, files read, write entropy, and file type coverage as file system activities \cite{ref_article27}. The various researchers recorded IRP open, IRP write, IRP create for IRP logs  \cite{ref_article26}. Chen et al.  \cite{ref_article29} generated API CFG (Call Flow graph) by API sequence extracted from the API monitor tool. They converted CFG into a feature vector for further processing.

\section{Implementation}

Our work has been carried out with a publicly available dataset prepared by Sgandurra et al.  \cite{ref_article24}. The dataset includes 582 ransomware samples and 942 goodware samples. Dataset authors have collected good ware and ransomware samples from various sources and used Cuckoo sandbox for analyzing them. Goodware samples include browsers, drivers, multimedia tools, file and network utilities, etc. Ransomware includes it's various families tabulated in Table \ref{tab1}.

\begin{table}
\begin{center}
\caption{Ransomware family names with correspondence ID.}\label{tab1}
\begin{tabular}{|l|l|}
\hline
\textbf{Family Name} & \textbf{ID} \\
\hline
Goodware & 0 \\
Critroni & 1 \\
CryptLocker & 2 \\
CryptoWall & 3 \\
KOLLAH & 4 \\
Kovter & 5 \\
Locker & 6 \\
MATSNU & 7 \\
PGPCODER & 8 \\
Reveton & 9 \\
TeslaCrypt & 10 \\
Trojan-Ransom & 11 \\
\hline
\end{tabular}
\end{center}
\end{table}

This dataset consists of a total 1524 records and 30970 features  \cite{ref_article24}. Features are set of various system events and activities, and they are identified with codes tabulated in Table \ref{tab2}. Feature values are in binary form, representing whether that event occurred during ransomware and goodware executables execution or not. If that event has occurred during execution, then the value of that feature column is 1 for that specific sample; otherwise, the value for that feature column is 0 for that specific sample. Like this, we have obtained a collection of binary values as a data frame for further analysis.

\begin{table}
\begin{center}
\caption{Features ID with description.}\label{tab2}
\begin{tabular}{|l|l|}
\hline
\textbf{ID} & \textbf{Description} \\
\hline
API & API invocations \\
DROP & Extensions of the dropped files \\
REG & Registry key operations \\
FILES & File operations \\
FILES\_EXT & Extension of the files involved in
file operations \\
DIR & File directory operations \\
STR & Embedded strings \\
\hline
\end{tabular}
\end{center}
\end{table}

Each column value of the dataset is mapped with several features shown in the above table, such as API invocation, registry key and files operations (write, read, create, delete), dropped file information, embedded strings, etc.

\subsection{Feature Selection}

\begin{figure}
\includegraphics[width=\textwidth]{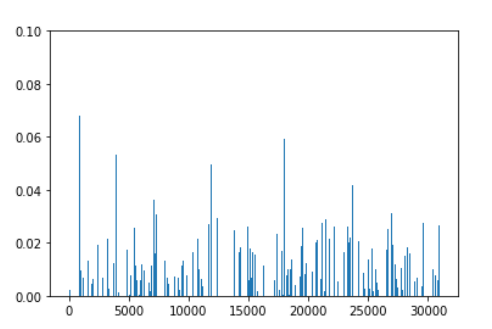}
\caption{MI score evaluated for each feature space.} \label{fig2}
\end{figure}

The next we have followed is feature selection. As we had 30970 features, we have implemented a Correlation-based Feature Selection (CFS) technique to select relevant features from a set of features. CFS evaluates the worth of a subset of attributes by considering each feature's predictive ability and the degree of redundancy between them. Subsets of features that are highly correlated with the class while having low intercorrelation are preferred. So, we decided to use the Mutual Information (MI) feature selection method. It is a more robust option for correlation estimation, which measures mutual dependence between variables. Mutual Information classification algorithm takes target features, and features set and evaluate a MI score for each feature in the set for the target feature. MI score shows how much that feature is dependent on the target feature. MI relies on entropy estimation from K-nearest neighbour’s distances. If the MI score of any feature equals zero, the feature is independent of the target feature. So, these features are none of our use, and we can neglect the same. If the MI score of any feature is greater than 0, it means that the feature is dependent on the target feature, and we can consider those features for our model. Higher the MI score, features are more dependent on the target feature, which is a good option. MI score resulted for our chosen dataset features are shown in Figure \ref{fig2} . Next, we have used the SelectKBest feature selection technique to select a set of top MI scored features, i.e., a set of most dependent features to provide input to the ML models.

\subsection{Model Implementation}

After selecting relevant features, we have moved forward towards model implementations. We have implemented various machine learning models on the selected feature set of chosen datasets, and those ML models are Decision Tree, Random Forest, KNN, SVM, XGBoost and Logistic Regression.

The decision tree is based on the concept of decision-making in each step. It forms a tree-like structure with possible outcomes, including resource cost, chance event consequences, and utility. This algorithm demonstrates that it contains conditional control statements. We have trained our decision tree model on our dataset. Next, we implemented ransom forest, a classification algorithm consisting of decision trees and uses feature randomness while building uncorrelated decision tree forest whose prediction is more accurate than any individual decision tree. The random forest has also given better results for our dataset. We have also implemented KNN (K-Nearest neighbour) classification method to classify whether an executable is goodware or ransomware. This method classifies data based on the concept of how similar two data points are. Test data is used to make an “educated guess”, like class, an unclassified data point should be classified. The goal of the SVM algorithm is to create the best line or decision boundary that can segregate n-dimensional space into classes so that we can quickly put the new data point in the correct category in the future. This best decision boundary is called a hyperplane. We trained our SVM model on our dataset and obtained good precision and recall; the result section discussed the same. Next, we have chosen XGBoost, an ensemble decision-tree-based mechanism based on a gradient boosting framework. When we implemented our dataset on XGBoost, we obtained the highest accuracy. At the same time, we have also implemented the same dataset with the Logistic regression (LR) machine learning model, which is used to allocate observation to a discrete set of classes. LR uses sigmoid function and produce a probability value, and based on the threshold on probability value; we used to decide in which any sample can be classified. The hypothesis of LR tends to limit the cost function in the range of 0 to 1.

\section{Results}

As several popular ML models have been implemented in this work, we have made a comparative analysis between these model’s efficiency. We have tested each implemented ML model on the test dataset and compared the results on various metrics, those are explained in detail below:

\begin{enumerate}
    \item \emph{Accuracy:} It means the fraction of predictions our model got right, i.e., the ratio of no of correct prediction to the total number of samples. Comparative accuracy analysis is shown in Table \ref{tab3}. \\
    \begin{equation}
        Accuracy = \frac{Number of Correct Prediction}{Total number of Prediction Made}
    \end{equation}
    
    \begin{table}
    \begin{center}
    \caption{Accuracy obtained by each model.}\label{tab3}
    \begin{tabular}{|l|l|}
    \hline
    \textbf{Model Name} & \textbf{Accuracy} \\
    \hline
    Decision Tree & 95.63 \\
    Random Forest & 96.02 \\
    KNN & 93.64 \\ 
    SVM &  96.42 \\ 
    XGBoost & 98.21 \\
    Linear Regression &  98.21 \\
    \hline
    \end{tabular}
    \end{center}
    \end{table}
    
    \item \emph{Confusion Matrix:} It provides a matrix as output and describes the complete performance of any ML model. For our case, we are solving binary class classification, and we had a total of 503 samples in our test data. There are four most relevant terms: (1) True Positive (TP) – The number of cases that are predicted as ransomware and the actual expected output is also ransomware only, (2) True Negative (TN) – The number of cases that are predicted as goodware and actual expected output is also goodware, (3) False Positive (FP) – The number of cases which are predicted as ransomware and expected output is goodware, (4) False Negative (FN) - The number of cases which are predicted as goodware and expected output is ransomware. The score of each of these terminologies are shown in Table \ref{tab4}.
    
    \begin{table}
    \begin{center}
    \caption{TP, TN, FP and FN obtained by each model.}\label{tab4}
    \begin{tabular}{|l|l|l|l|l|}
    \hline
    \textbf{Name} & \textbf{TP} & \textbf{TN} & \textbf{FP} & \textbf{FN} \\
    \hline
    DT & 184 & 297 & 16 & 6 \\
    RF & 186 & 297 & 16 & 4 \\ 
    KNN & 181 & 299 & 23 & 9 \\
    SVM & 185 & 300 & 13 & 5 \\
    XGB & 188 & 306 & 7 & 2 \\ 
    LR & 187 & 307 & 6 & 3 \\
    \hline
    \end{tabular}
    \end{center}
    \end{table}
    
    \item \emph{Precision \& Recall:} Precision shows the number of correct positive outputs in total positive results predicted by the classifier. Recall demonstrates no of correct positive outputs in all samples that should have been resulted as positive. Comparative analysis is shown in Table \ref{tab5}.
    \begin{equation}
        Precision = \frac{TP}{TP+FP}
    \end{equation}
    \begin{equation}
        Recall = \frac{TP}{TP+FN}
    \end{equation}
    
    \begin{table}
    \begin{center}
    \caption{Precision and Recall obtained by each model.}\label{tab5}
    \begin{tabular}{|l|l|l|}
    \hline
    \textbf{Name} & \textbf{Precision} & \textbf{Recall}\\
    \hline
    DT & 0.92 & 0.97 \\
    RF & 0.92 & 0.98 \\
    KNN & 0.89 & 0.95 \\
    SVM & 0.93 &  0.97 \\
    XGB & 0.96 & 0.99 \\
    LR &  0.97 & 0.98 \\
    \hline
    \end{tabular}
    \end{center}
    \end{table}
    
\end{enumerate}

\section{Conclusion \& Future Works}

This work aimed to design and develop an ML-based system that identifies executables whose activity matches ransomware activities. This thesis works upon working on well-defined behaviour, and the intention of executables are being analyzed. This framework consists of steps: Data collection, processing of those data, features extraction and implementation of ML model. The same has been achieved by deploying Cuckoo Sandbox, Ransomware sample analysis and ML model implementation using python script.

The implemented ML model can get extended to implement an ML-based detection system sole product to detect upcoming ransomware threats for the industry. For enterprise-level implementation, we need to monitor logs generated by the organization’s distributed systems. We can feed the status of API calls log, registry key operations logs, and file system operations logs to our saved model to decide whether the activity belongs to ransomware or not.

To record these logs activities, enterprises generally use EDR (Endpoint Detection and Response) and SIEM (Security Information and Event Management). EDR is an endpoint security solution that combines endpoint data and real-time continuous log monitoring with a rule-based automated response. SIEM is a security tool, which collects data(logs) from devices, domain controllers, networks, servers. It stores those data and applies analytics on stored data to detect threats, discovers trends, and enables organizations to examine any alerts. Several EDR and SIEM solutions are available in the industry, and we can integrate this ML model with any of these solutions to detect ransomware activities.

Also, we can use ELK (Elasticsearch Logstash Kibana) \cite{ref_article30} to integrate our ML model and develop a ransomware detection system as a product. ELK is an open-source tool, and many SIEMs are built on top of ELK only. We can use ELK for elastic search to filter and collected desired logs for our detection system. The implemented model can get ingested the collected logs to decide on suspicious activity based on logs details. ELK stack is a collection of three open-source tools, i.e. Elasticsearch, Logstash, Kibana, which allow us to store several logs to visualize them easier. Elasticsearch is a search and analysis engine that allows us to store massive information in JSON format. Logstash is a data processor and collector, which allows us to collect, load, and transfer data in numerous architectures, and in the last, it can get forwarded to Elasticsearch. Kibana provides a visualization layer to a web browser to parse any query to Elasticsearch in several visualizations like histograms, linear graphs, pie charts, etc. We can use ELK directly to integrate our model with it as ELK is an open-source project.

%
%
%
%

\end{document}